\def\Anonym{0} % anonymize?
\def\Draft{0} % show comments (todos), ToC, etc.
\def\Class{2} % 0:lncs 1:iacr 2:ieee 3:lipics

\if\Class0
	\documentclass{styles/llncs}
    \usepackage{geometry}
\fi
\if\Class1
    \def\IacrName{tches}
	\documentclass[journal=\IacrName,preprint,spthm]{styles/iacrtrans}
\fi
\if\Class2
    \documentclass[conference]{styles/IEEEtran}
    \IEEEoverridecommandlockouts
\fi
\if\Class3
    \documentclass[a4paper,UKenglish,cleveref, autoref, thm-restate]{styles/lipics}
\fi

\usepackage[misc,geometry]{ifsym} % To have the letter symbol
\usepackage{stfloats}
\usepackage{placeins}

\usepackage{graphicx}
\usepackage{url}
\PassOptionsToPackage{unicode,hidelinks,colorlinks=true,backref=page}{hyperref}
\usepackage{hyperref}

\if\Class2
\else
\usepackage{appendix}
\fi

\usepackage{cite}
\usepackage{xifthen}
\usepackage{import}
\usepackage{bookmark} % disables some warnings
\usepackage{float}
\usepackage{booktabs}
\usepackage{multirow}
\usepackage{siunitx}
\usepackage{comment}
\usepackage{makecell}
\usepackage{algorithm}
\usepackage{algorithmic}
\usepackage{tabularx}

\if\Class0
\let\proof\relax

\fi

\usepackage{amsmath}
\usepackage{amsthm}
\usepackage{amssymb}
\usepackage{mathtools} % \coloneqq
\mathtoolsset{showonlyrefs}

\usepackage{extarrows}  % more arrows under text / over text, xlongarrow 

\usepackage{xspace} % add \xspace to end of \newcommand to avoid putting {} in each call

\usepackage{tikz}
\usetikzlibrary{arrows.meta, positioning}
\usepackage{fontawesome5}

\usetikzlibrary{arrows}
\usetikzlibrary{patterns}
\usetikzlibrary{shapes}
\usetikzlibrary{calc}
\usetikzlibrary{decorations.pathreplacing}
\if\Class3
\else
\usepackage{paralist} % e.g. \begin{enumerate}[(i)] for roman enum
\fi
\if\Class2
\fi
\usepackage[colorinlistoftodos,textsize=small]{todonotes}

\usepackage[strings]{underscore} % fix doi in bib (splncs04 issue)

\usepackage{tikz}
\usetikzlibrary{shapes,calc,positioning,arrows,}
\pgfarrowsdeclarecombine{twotriang}{twotriang}%    double headed arrow
{stealth'}{stealth'}{stealth'}{stealth'}

\usepackage{cases}
\usepackage{mathtools}

\graphicspath{{figures/}}

\newcommand\pround[1]{\left( #1 \right)}

\newcommand\ifPar[1]{%
	\ifthenelse{\isempty{#1}}%
		{}%
		{\pround{#1}}%
}
\newcommand\ifParElse[2]{
	\ifthenelse{\isempty{#1}}{#2}{(#1)}
}

\title{Towards Cost-Effective ZK-Rollups: Modeling and Optimization of Proving Infrastructure
}

\if\Class2
\else
\titlerunning{Towards Cost-Effective ZK-Rollups}
\fi

\if\Class3
  \if\Anonym0
    \author{Mohsen {Ahmadvand}}{Zircuit}{name@zircuit.com}{}{}
    \author{Pedro {Souto}}{Zircuit}{name@zircuit.com}{}{}
    \Copyright{Zircuit}
    \authorrunning{Ahmadvand, Vitto and Souto}
  \else
    \author{}{}{}{}{}
    \Copyright{}
  \fi
\fi

\if\Class2
  \if\Anonym0
    \author{
      \IEEEauthorblockN{Mohsen Ahmadvand and Pedro Souto}
      \IEEEauthorblockA{Zircuit\\
      \texttt{\{name\}@zircuit.com}}
    }
  \else
  \fi
\fi

\if\Class1
  \if\Anonym0
    \author{
      Mohsen Ahmadvand \and Pedro Souto
    }
    \institute{
      Zircuit \\
      \email{{name}@zircuit.com}
    }
  \else
    \author{}
    \institute{}
  \fi
\fi

\if\Class0
  \if\Anonym0
    \author{
      Mohsen Ahmadvand \and Pedro Souto
    }
    \institute{
      Zircuit \\
      \email{{name}@zircuit.com}
    }
  \else
    \author{}
    \institute{}
  \fi
\fi

\if\Draft1
\newcommand\MA[1]{\todo[author=MA,inline,color=red!20!white]{#1}}
\newcommand\GV[1]{\todo[author=GV,inline,color=blue!20!white]{#1}}
\newcommand\PS[1]{\todo[author=PS,inline,color=green!20!white]{#1}}
\else
\newcommand\MA[1]{}
\newcommand\GV[1]{}
\newcommand\PS[1]{}
\newcommand\MAdone[1]{}
\newcommand\GVdone[1]{}
\newcommand\PSdone[1]{}
\fi

\begin{document}

    \renewcommand{\sectionautorefname}{Section}
    \renewcommand{\subsectionautorefname}{Subsection}
    \newcommand{\aref}[1]{\hyperref[#1]{Appendix~\ref*{#1}}}

    \maketitle

    \if\Class0
    \setcounter{tocdepth}{10} % because of lncs
    \fi
    
    \if\Draft1
    \listoftodos
    \tableofcontents
    \fi
    
    % Main body
    % Keywords
\newcommand\kwsnoand{
Zero-Knowledge, Rollups, Provers, Cost Modeling, SMT Solver
}
\newcommand\kwsand{
Zero-Knowledge \and Rollups  \and Provers  \and  Cost Modeling  \and  SMT Solver
}

\if\Class3
\fi

\if\Class2
\fi

\if\Class1
\fi

\begin{abstract}
\if\Class0
\fi
Zero-knowledge rollups rely on provers to generate multi-step state transition proofs under strict finality and availability constraints. These steps require expensive hardware (e.g., GPUs), and finality is reached only once all stages complete and results are posted on-chain. As rollups scale, staying economically viable becomes increasingly difficult due to rising throughput, fast finality demands, volatile gas prices, and dynamic resource needs.

We base our study on Halo2-based proving systems and identify transactions per second (TPS), average gas usage, and finality time as key cost drivers. To address this, we propose a parametric cost model that captures rollup-specific constraints and ensures provers can keep up with incoming transaction load. We formulate this model as a constraint system and solve it using the Z3 SMT solver to find cost-optimal configurations. To validate our approach, we implement a simulator that detects lag and estimates operational costs. Our method shows a potential cost reduction of up to 70\%.

\end{abstract}
\section{Introduction}

Rollups are specialized blockchains designed to improve scaling or introduce new features to existing blockchains~\cite{gorzny2024rollup}. They currently hold \$55B USD in value locked, playing a significant role in the ecosystem~\cite{l2beat_scaling_summary}.
To use rollups, users typically send cryptocurrencies to a \textit{bridge} on the main chain (e.g., Ethereum) and receive an equivalent amount on the rollup chain~\cite{mccorry2021sok}. After transacting, users can \textit{exit} the chain by bridging back to the main chain, also known as \textit{withdrawal}~\cite{gorzny2022ideal}. Transactions on the rollup result in state transitions, which are summarized and stored on the main chain in a process called finalization~\cite{optimism_specs}. Finalization involves submitting data availability and proof verification to the main chain, incurring what are known as L1 fees.

ZK-rollups generate a proof of execution, commonly referred to as Zero-Knowledge Succinct Non-Interactive Argument of Knowledge (zk-SNARK)~\cite{ghadafi2010groth}, computed by a prover. This proof is verified, for example, on the main chain to finalize state transitions, enabling shorter wait times and trustless operations.

The top three fees paid by chains—Scroll, ZKSync, and Linea—over a year amount to \$21.29M, \$16.69M, and \$12.01M USD, respectively~\cite{l2beat_onchain_costs}. These figures represent only the on-chain fees paid on the Ethereum network (L1 fees), while a significant hidden layer of costs, namely prover running expenses, remains unaccounted for. The cost efficiency of these numbers is difficult to assess due to the limited availability of detailed information.

With such high expenses, ZK-rollups face the risk of becoming non-viable. A rollup remains viable only if the total cost per transaction stays significantly below L1 fees while maintaining competitiveness within the ecosystem.

Three key metrics in rollup design are transactions per second (TPS), average gas usage per transaction ($Gas_{PerTx}$), and target finality (\(T_{Finality}\)). These metrics define the system's performance and target capacity: the maximum TPS indicates the rollup's throughput capacity, $Gas_{PerTx}$ caputures the expected average gas usage of transactions, while the finality upper-bound provides assurance to users and exchanges on the time frame within which a certain state of the chain will be finalized. All metrics have a direct impact on the configuration of provers and their resource requirements. Supporting higher TPS or $Gas_{PerTx}$ necessitates greater resource availability, while lower finality targets require a higher rate of submissions to the main chain, thereby increasing base layer (L1) fees.

Running provers is often underestimated, with many of the operational challenges hidden beneath the surface. Proving state transitions requires a sequence of tightly coordinated parallel and sequential steps, each of which must execute reliably. This process is highly resource-intensive, demands specialized hardware, and has limited throughput per proving cycle~\cite{ma2023gzkp}.

A failure in the proving process can halt the rollup entirely—a condition known as \textit{chainhalt}—in which users are unable to withdraw their funds~\cite{gorzny2022ideal}. Such delays undermine trust, damage the rollup’s reputation, and may break integrations with applications that rely on deterministic finality.

On average, proving time scales with the number of transactions that gets validated. When, due to an higher chain traffic, the prover needs to validate more transactions than usual, users will then experience longer transaction delays, resulting in \textit{delayed finality}. Furthermore, if the allocated resources are incapable to scale in order to handle such extra load, the excess traffic will accumulate, creating a compounding situation where the rollup latest finalized state lags further behind as time passes. 

The challenge lies in reliably identifying the optimal configuration for the setup, as over-allocating resources increases costs, potentially rendering the rollup financially unviable.

Since there are multiple steps that need to synchronize and advance in a harmonic manner, it is crucial to allocate resources in a balanced way, ensuring that all steps have sufficient resources, remain cost-efficient, and conform to the system's target TPS and finality. TPS can fluctuate during system operation; for instance, the system may experience excessive load during a popular airdrop.

\textbf{Problem.}  
Proving pipelines have become increasingly complex, with interdependent steps and strict system constraints such as finality. The core challenge is maintaining economic viability while adhering to these constraints. Costs are difficult to predict and optimize due to multiple factors, including base layer fees, hardware scaling, and fluctuating inbound transaction rates.

\textbf{Contributions.}  
To address the above challenges, we:  
\begin{enumerate}[(i)]
    \item propose a cost-estimation model that incorporates throughput (TPS), average gas per transaction, and finality constraints to enable lag-free prover configuration and accurate cost prediction;
    \item formulate the cost model as a constraint system and solve it using the Z3 SMT solver to obtain cost-optimal prover configurations;
    \item implement a simulator to evaluate prover setups, detect lag, and estimate operational costs under varying conditions;
    \item applied our optimization to four real-world scenarios based on the public data from the Zircuit chain, demonstrating a potential cost reduction of up to 71\%.
\end{enumerate}

This paper addresses critical zk-rollup proving infrastructure cost optimization, an area often overlooked despite its essential role. By focusing on the hidden complexities of proving pipelines, it provides insights and practical solutions that extend beyond simple zkEVM circuits and Geth configurations.

\section{Preliminaries}

\subsection{The Halo2 Proving System}
A zero-knowledge proof system enables a prover to convince a verifier that a certain statement is true without revealing any information beyond the validity of the statement itself. Formally, the statement is encoded as membership in a language, where the language comprises a set of valid instances defined by a relation. A \emph{witness} is the secret input demonstrating that the instance satisfies the relation. Zero-knowledge proofs produce succinct cryptographic proofs attesting that a valid witness exists for the given statement, while ensuring soundness—that is, it is computationally infeasible for any adversary to produce a valid proof without knowledge of a valid witness—and zero-knowledge, guaranteeing no additional information about the witness leaks.

In the Halo2 proof system, statements are represented as arithmetic circuits composed of custom gates. These gates generalize Boolean logic gates and impose polynomial constraints over numerical values assigned to circuit wires. The constraints are organized within a structured constraint system implemented as a table, partitioned into columns that serve different roles—such as \emph{advice columns} (holding witness-assigned values), \emph{fixed columns} (containing constant or selector values), and \emph{lookup columns} (used for efficient range checks or value lookups). The semantics and interaction of these column types define the circuit’s behavior during \emph{circuit synthesis}, the process where the prover assigns witness values subject to all constraints.

Given a fixed number of columns, the capacity of a Halo2 circuit is primarily limited by the number of rows in the constraint table, which determines the total number of values that can be processed. For instance, to verify that all elements in an input vector satisfy certain properties (e.g., being even), the circuit requires a proportional allocation of rows per element. Consequently, the maximum input size or complexity the circuit can handle is bounded by the total number of rows available.

\subsubsection{Row Limits and Circuit Capacity}
The number of rows available in the constraint system depends on the proving system parameters. In particular, for the KZG polynomial commitment scheme employed in this work, the maximum number of rows is bounded by $2^{28}$. Increasing the number of rows increases prover computation time and memory usage, thereby impacting the overall proof generation cost. Hence, selecting an optimal number of rows—balanced against the input size and required computations—is critical for efficient proof construction.

The row limit acts as a hard upper bound on the circuit’s capacity to handle inputs. To enforce this, a \emph{circuit capacity checker (CCC)} validates that the input size does not exceed the allowed maximum number of rows. This check prevents attempts to generate proofs for statements that exceed the circuit’s predefined capacity.

\subsubsection{Recursive Proofs}
Recursion enables the compression of multiple proofs into a single aggregated proof by verifying constituent proofs within a circuit itself. Recursive proof composition allows one to produce a succinct proof attesting to the validity of a set of proofs, thereby reducing verification complexity and proof size in aggregate. This technique is particularly valuable for scalable proof systems and is supported by Halo2’s flexible circuit design.

\subsection{zkEVM Circuits}

The zkEVM Circuits \cite{pse_zkevm_circuits} are a collection of Halo2-based circuits that collectively capture the semantics of the Ethereum Virtual Machine (EVM) by encoding the constraints necessary to verify the validity of Ethereum transactions. The architecture of the zkEVM is modular, consisting of specialized sub-circuits, each responsible for enforcing a specific aspect of EVM semantics. For instance, dedicated circuits handle transaction validation, EVM bytecode execution, state transition logic, Keccak hashing, and Merkle Patricia Trie (MPT) proof verification. By embedding EVM correctness directly into these circuits through custom gate constraints, the zkEVM ensures that only valid state transitions—provided as witness data—can be proven and subsequently verified. 

Since different sub-circuits are responsible for validating different components of an Ethereum block, proving the validity of an entire block requires a unified approach. This is achieved through the {\it SuperCircuit}, a higher-level circuit that aggregates all sub-circuits into a single construct. The SuperCircuit eliminates redundant data replication across sub-circuits where possible and ensures consistent constraint enforcement across the full scope of the EVM state transition. A proof generated by the SuperCircuit attests to the validity of the entire block as a whole, rather than relying on multiple separate proofs for each individual component.

In practice, the SuperCircuit can also process multiple blocks within a single proof by leveraging available rows in the constraint system. When blocks contain fewer transactions or require fewer resources, multiple such blocks may be aggregated and proved together, maximizing circuit capacity utilization. Conversely, for blocks with high transaction density that approach the circuit’s row limit, the SuperCircuit may prove only one block at a time. This tradeoff is managed by the \emph{circuit capacity checker (CCC)}, which ensures that the input size does not exceed the maximum number of rows allowed by the underlying KZG commitment scheme parameters.

\subsection{Transactions Per Second (TPS) and Average Gas Usage per Transaction}

Transactions Per Second (TPS) is a fundamental performance metric for any blockchain or rollup system, representing the throughput capacity—the maximum number of transactions the system can process per second. The average gas usage per transaction (GasPerTx) measures the average computational and storage cost of a transaction on Layer 1 (Ethereum), directly impacting the fees required for data availability and proof verification.

Both TPS and GasPerTx serve as fixed parameters in our cost model and influence the design of proving infrastructure and on-chain costs. Higher TPS demands necessitate more prover resources to handle increased transaction volume, while higher GasPerTx increases Layer 1 costs due to greater data and proof complexity.
\subsection{Zero-Knowledge Rollup}

Zero-Knowledge Rollups (ZK-rollups) are Layer 2 scaling solutions that aggregate multiple transactions off-chain and submit succinct zero-knowledge proofs to the Ethereum mainnet to attest to the validity of state transitions. ZK-rollups reduce on-chain gas costs and improve throughput by compressing transaction data and proofs into small, verifiable proofs.

Several critical parameters influence the cost and performance of ZK-rollups:

\subsubsection{Transaction Per Second (TPS)}
TPS is a fixed parameter that defines the target throughput for the rollup. The prover infrastructure must scale to meet this throughput, ensuring that the rollup can process transactions as they arrive without delay.

\subsubsection{Finality}
Finality refers to the maximum acceptable time to finalize a batch of transactions on Layer 1. A shorter finality target increases system responsiveness but generally leads to higher costs, as fewer transactions can be aggregated per batch, resulting in more frequent submissions and associated Layer 1 fees.

\subsubsection{Data Availability}
ZK-rollups must submit transaction data or proofs on-chain for data availability. This submission incurs gas costs proportional to the data size and submission method. There are two primary models:

\emph{Calldata model:} The rollup pays gas fees per byte of calldata submitted on-chain.

\emph{Blob model:} The rollup pays a fixed fee per blob, regardless of the blob’s actual utilization.

Gas prices on Ethereum fluctuate dynamically, affecting overall cost. The data availability cost is therefore modeled as a product of a fixed gas fee parameter and the size of data submitted.

\subsubsection{Proof Verification}
Proof verification on Layer 1 incurs a fixed gas cost per proof bundle, which must be factored into total Layer 1 fees.

\subsubsection{Prover Infrastructure}
The prover is the off-chain component responsible for generating zero-knowledge proofs. Our analysis focuses on Halo2, a battle-tested proof system originally developed by Zcash and employed by projects such as Privacy and Scaling Explorations (PSE), Scroll, Taiko, Zircuit, and Koroma. The prover system must be sized and optimized to keep up with the target TPS while respecting finality constraints.

\paragraph{Importance of Target TPS and Finality}
The upper-bound or target TPS is crucial because the system must achieve finality within a defined time frame given this transaction load. Prover resources and batch sizes must be configured such that proofs are generated and verified promptly to prevent delays in finalization, which would degrade user experience and trust.

\subsection{Proving Pipelines}

A ZK-rollup must generate proofs according to its proof system and circuit composition. In the system under study, super circuit proofs are executed fully in parallel, without synchronization. As transactions arrive, they are packed and dispatched to super circuit provers. To handle higher throughput, more super provers must be commissioned dynamically, as each prover has a fixed capacity per proving cycle.

These super proofs are aggregated into batch proofs, each comprising a fixed number of sequential super proofs. The batcher operates with a configurable epoch and may introduce delay to accumulate enough proofs before proceeding. If the number of super proofs exceeds the capacity of a single batch prover, additional batch provers must be commissioned to sustain throughput. Smaller batches enable more frequent submissions but increase data availability (DA) costs, as each batch requires posting calldata or blob data on-chain.

Batch proofs are further aggregated into bundle proofs via recursive composition. Bundle provers select from available batch proofs based on target finality and system conditions. Smaller bundles reduce latency but increase L1 verification costs, since each bundle must be verified individually on-chain.

The proving pipeline must balance prover parallelism, batching delays, and recursion depth to meet finality constraints while remaining economically viable. Misconfiguration can result in prover backlogs, underutilized hardware, or excessive on-chain costs.

% \usepackage{tikz}
% \usetikzlibrary{arrows.meta, positioning}

\begin{figure}[t]
\centering
\begin{tikzpicture}[
  node distance=0.6cm and 0.9cm,
  box/.style={draw, rectangle, rounded corners, minimum width=2cm, minimum height=0.7cm, font=\scriptsize, align=center},
  arrow/.style={-{Latex}, thick},
  font=\scriptsize
]

% Main nodes
\node[box] (txs) {Inbound TXs\\Subject to TPS\\ and $Gas_{PerTx}$};
\node[box, below left=of txs, xshift=-0.3cm] (super1) {Super\\Prover 1};
\node[box, below right=of txs, xshift=0.3cm] (super2) {Super\\Prover N};
\node[box, below=1.3cm of txs] (batch) {Batch\\Prover(s)};
\node[box, below=0.8cm of batch] (bundle) {Bundle\\Prover};
\node[box, right=0.8cm of bundle] (da) {DA per\\Batch};
\node[box, below=0.8cm of bundle] (l1) {L1\\Verify Per Bundle};

% Arrows between stages
\draw[arrow] (txs) -- (super1);
\draw[arrow] (txs) -- (super2);
\draw[arrow] (super1) -- (batch);
\draw[arrow] (super2) -- (batch);
\draw[arrow] (batch) -- (bundle);
\draw[arrow] (bundle) -- (l1);

% Arrow to DA box (parallel to bundle)
\draw[arrow] (batch.east) -- (da.west);

% Top annotations (TPS scaling and batch scaling)
\node[font=\scriptsize, above=0.35cm of super1, align=center] (scale1) {↑ scales with TPS\\\faClock~Wait for enough TXs};

\node[font=\scriptsize, left=0.35cm of bundle, align=center] (bundlenote) {\faClock~Wait for enough batch\\ proofs before publishing};
\draw[arrow] (bundlenote.east) -- (bundle.west);

\draw[arrow] (scale1.south) -- (super1.north);

\node[font=\scriptsize, above=0.35cm of batch, align=center] (scale2) {↑ scale with Super provers\\Wait for enough super proofs};
\draw[arrow] (scale2.south) -- (batch.north);

\end{tikzpicture}
\caption{Proving pipeline: inbound transactions trigger super provers; outputs are batched and recursively bundled. Super prover count scales with TPS; excess proofs may require more batch provers. Each batch incurs DA cost; each bundle is verified on L1.}
\label{fig:proving-pipeline}
\end{figure}
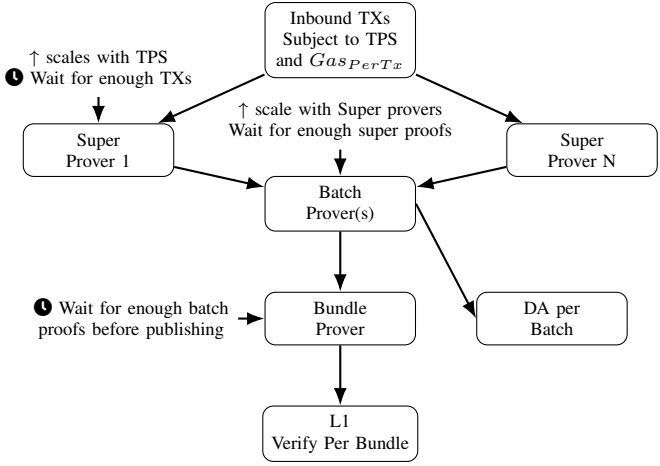

\subsection{Costs}

We distinguish between \emph{fee}, the amount paid by users, and \emph{price}, the cost incurred by the rollup.

\subsubsection{Hardware Price}
Fixed monthly cost parameter. When purchased, costs can be amortized over a number of years.

\subsubsection{Layer 1 Price}
Includes fixed gas costs for proof verification and data availability, scaled by gas price and data size.

\subsubsection{Total Price}
Dynamic per-transaction cost combining hardware and Layer 1 expenses, depending on throughput, gas usage, and finality targets.

\section{Design}
As discussed earlier three parameters that predominantly influence these costs are \textit{Transactions Per Second (TPS)}, $Gas_{PerTx}$, and \textit{Target Finality}. By focusing on these parameters, our model provides a framework for balancing resource allocation and scheduling to minimize costs while maintaining the system's operational goals.

To build the infrastructure required for running the provers in the proposed ZK-rollup, we need to set fixed values derived from runtime benchmarks, such as the price per machine, maximum memory, and execution time. Additionally, constant values from the setup, such as gas fees for data availability (DA) and verifications, as well as circuit maximum capacity, must also be defined.

There are two inherently dynamic variables: the L1 gas price and the ETH price in USD. We assume these values are fixed for each iteration of the cost estimation. Running the estimation with updated values provides the flexibility to analyze the impact of fluctuations in these variables.

Once the fixed parameters are configured, the model computes the corresponding configuration parameters. The goal is to meet the TPS requirements and finality targets at the lowest possible cost based on the fixed parameters.

\subsection{Fixed Parameters}\label{sec:fixed-params}
Below we listed the fixed parameters that drive cost estimations:
\begin{itemize}
    \item \textbf{Average Gas Per Tx.} The average gas used per Ethereum transaction typically ranges from 80k to 120k, depending on network conditions and transaction complexity. We adopt 100k as a representative average, consistent with data from Etherscan and L2BEAT.
    \item \textbf{Total Block Gas.} 10m is the hardcoded value we use on our chain under study.
    \item \textbf{TPS:} Target transactions per second (can fluctuate).
    \item \textbf{Target Finality (\(T_{\text{finality}}\)):} The system's finality target.
    \item \textbf{Max TXs in Super Circuit (\(\text{Tx}_{\text{max}}\)):} Maximum transactions processed in the super circuit.
    \item \textbf{Max Super Proofs in Batch Circuit (\(\text{Max}_{\text{super proofs per batch}}\)):} Maximum super proofs processed in the batch circuit.
    \item \textbf{Max Batch Proofs in Bundle Circuit (\(\text{Max}_{\text{batch proofs per bundle}}\)):} Maximum batch proofs processed in the bundle circuit.
    \item \textbf{Hardware Cost per Machine per Month (\(\text{Cost}_{\text{hardware}}\)):} Monthly cost of provisioning hardware for proving.
    \item \textbf{Super Circuit Proof Generation Time (\(T_{\text{super}}\)):} Time taken to generate a proof in the super circuit per machine.
    \item \textbf{Batch Circuit Proof Generation Time (\(T_{\text{batch}}\)):} Time taken to generate a proof in the batch circuit per machine.
    \item \textbf{Bundle Circuit Proof Generation Time (\(T_{\text{bundle}}\)):} Time taken to generate a proof in the bundle circuit per machine.
    \item \textbf{L1 Proof Verification Gas Price (\(\text{Gas}_{\text{verification per bundle}}\)):} Gas cost for proof verification on L1.
    \item \textbf{L1 DA Submission Gas Price (\(\text{Gas}_{\text{DA per batch}}\)):} Gas cost for data availability submission on L1.
    \item \textbf{Ethereum Gas Price (\(\text{Gas}_{\text{price (Gwei)}}\)):} Gas price on Ethereum (fluctuates).
    \item \textbf{ETH Price in USD (\(\text{Price}_{\text{ETH/USD}}\)):} Price of Ethereum in USD (fluctuates).
\end{itemize}

To be able to estimate the cost and set the fixed parameters (\(T_{\text{super}}\), \(T_{\text{batch}}\), and \(T_{\text{bundle}}\)), we require real numbers for prover times and machine costs. First, we need to determine the memory requirements for each prover step to identify suitable machines that can handle the workload. We identify the memory requirements and subsequently suitable machines in~\autoref{tab:memory_usage_gb}.

\subsection{Derived Parameters}
Now, for the set fixed parameters, we aim to determine configurations that fulfill the TPS and target finality requirements while adhering to all other constants. We refer to these configurations as \textit{derived parameters}.
\begin{itemize}
    \item \textbf{Number of Super Circuit Provers (\(N_{\text{super}}\)):} The number of machines allocated for generating super proofs.
    \item \textbf{Number of Batch Provers (\(N_{\text{batch}}\)):} The number of machines allocated for processing batch proofs.
    \item \textbf{Number of Bundle Provers (\(N_{\text{bundle}}\)):} The number of machines allocated for processing bundle proofs.
    \item \textbf{Batch Epoch Time (\(\text{Epoch}_{\text{batch}}\)):} The time required to accumulate enough super proofs for a batch.
    \item \textbf{Bundle Epoch Time (\(\text{Epoch}_{\text{bundle}}\)):} The time required to accumulate enough batch proofs for a bundle.
    \item \textbf{Total Hardware Cost per Month (\(\text{Cost}_{\text{hardware total}}\)):} The cumulative cost of all prover machines per month.
    \item \textbf{Total L1 Cost per Month (\(\text{Cost}_{\text{L1 total}}\)):} The total monthly cost of L1 operations, including:
    \begin{itemize}
        \item \textbf{L1 Verification Costs (\(\text{Cost}_{\text{L1 verification}}\)):} Gas fees for proof verification on L1.
        \item \textbf{L1 Data Availability Costs (\(\text{Cost}_{\text{L1 DA}}\)):} Gas fees for data availability submissions on L1.
    \end{itemize}
\end{itemize}

\subsection{Target Finality}
To identify a configuration with minimal costs, it is crucial to consider finality requirements. Without incorporating finality needs, optimization may lead to configurations with excessively long finality times, especially at low TPS rates when the chain does not collect substantial gas fees. In such scenarios, finality times could extend to several days, which is generally unacceptable as the industry standard is significantly lower. 

To address this issue, it is necessary to allocate additional resources for both L1 costs and machine costs, enabling more frequent submissions to reduce finality times to an acceptable target. This adjustment ensures the rollup remains competitive and meets industry standards for finality.

At the time of writing, finality times were obtained from publicly available documentation \cite{scrollDocs}, \cite{polygonDocs}, and \cite{l2beatTaiko}, representing the current values listed in \autoref{tab:rollup-finality}. However, we were unable to find information about Kroma's finality time in any publicly available source.

\begin{table}[htbp]
\caption{Finality Times of Selected ZK-Rollups (in seconds)}
\centering
\begin{tabular}{|l|c|}
\hline
\textbf{Rollup} & \textbf{Finality Time (seconds)} \\ \hline
Scroll          & 3600                             \\ \hline
Polygon         & 3600                             \\ \hline
Taiko           & 14400                            \\ \hline
% Add additional rollups here as needed.
\end{tabular}
\label{tab:rollup-finality}
\end{table}

As per our finality time for the benchmarks, we select 4 hours (14400 seconds) as a suitable starting point for chains with low TPS load. This value is configurable in our parameters and can be adjusted to accommodate different requirements.

\subsection{Modeling the problem}
We aim to set the derived parameters to values that yield the lowest costs while conforming to a given target or latest finality. 

To achieve this, we model the problem as a set of constraints that ensure both TPS and finality are always guaranteed—in other words, the system never lags behind. Using this model, we compute the derived parameters as follows:

\subsubsection{Number of Super Provers}
\[
N_{\text{Super}} = \lceil\text{TPS} \cdot T_{\text{super}} / \text{Tx}_{\max}\rceil
\]
This computes the number of super provers required to handle the workload. Here, \(\text{Tx}_{\max}\) is the maximum number of transactions one super prover can handle in time \(T_{\text{super}}\).

\subsubsection{Batch Epoch}
The batch epoch \(\text{Epoch}_{\text{batch}}\) represents the time required to accumulate enough super proofs for batch provers. The \emph{unconstrained} batch epoch is determined by taking the maximum among \(T_{\text{super}}\), \(T_{\text{batch}}\), and the time required to distribute the workload among super provers. Mathematically, this can be captured as:
\[
\max\!\Bigg(
\,T_{\text{super}}, 
\,T_{\text{batch}}, 
\left\lceil 
\dfrac{\text{Max}_{\text{super proofs per batch}} \,\cdot\, T_{\text{super}}}
{N_{\text{super}}} 
\right\rceil
\Bigg).
\]

However, to conform to the target finality \(\bigl(T_{\text{finality}}\bigr)\), the batch epoch is capped to ensure that the total proving and bundling time does not exceed the finality constraint. Thus, we write:
\begin{multline}
\text{Epoch}_{\text{batch}} 
= 
\min\!\Bigg(\,
\max\!\Bigg(
T_{\text{super}}, 
\,T_{\text{batch}}, \\
\,\left\lceil 
\dfrac{
\text{Max}_{\text{super proofs per batch}} 
\,\cdot\, 
T_{\text{super}}
}
{N_{\text{super}}} 
\right\rceil
\Bigg), \\
T_{\text{finality}} 
- 
T_{\text{super}} 
- 
T_{\text{bundle}} 
\Bigg).
\end{multline}

\subsubsection{Number of Batch Provers}
To compute the number of batch provers required, we take into account the number of super proofs generated per epoch and the batching capacity of a single prover. The capacity of a batch prover is determined by how many batch proofs it can produce within the batch epoch.

\begin{equation}
N_{\text{batch}} 
= 
\left\lceil 
\dfrac{
N_{\text{super proofs per epoch}}
}{
\text{Max}_{\text{super proofs per batch}}
\,\cdot\,
\left\lfloor 
\dfrac{
\text{Batch Epoch}
}{
T_{\text{batch}}
}
\right\rfloor
}
\right\rceil
\end{equation}

\paragraph*{Explanation}
\begin{itemize}
    \item \(N_{\text{super proofs per epoch}}\): The total number of super proofs generated during one batch epoch.
    \item \(\text{Max}_{\text{super proofs per batch}}\): The maximum number of super proofs that can be processed in a single batch.
    \item \(\text{Batch Epoch}\): The time duration of one batch epoch.
    \item \(T_{\text{batch}}\): The time it takes for a single batch prover to process one batch.
    \item \(\left\lfloor \frac{\text{Batch Epoch}}{T_{\text{batch}}} \right\rfloor\): The number of batches a single prover can process in one epoch.
\end{itemize}

This formula ensures that the number of batch provers is computed based on the effective batching capacity of a single prover within the epoch, thus avoiding overestimation. By accounting for the batch epoch and processing time, we optimize the number of provers required to handle the load efficiently.

\subsubsection{Bundle Epoch}
The bundle epoch \(\text{Epoch}_{\text{bundle}}\) represents the time required to accumulate enough batch proofs to form a bundle. The \emph{unconstrained} bundle epoch is initially determined by the maximum between \(\text{Epoch}_{\text{batch}}\) and the time needed to distribute the workload across batch provers:
\[
\max\!\Biggl(\text{Epoch}_{\text{batch}}, 
\;\left\lceil 
\dfrac{
\text{Max}_{\text{batch proofs per bundle}}
\,\cdot\,
\text{Epoch}_{\text{batch}}
}{
N_{\text{batch}}
} 
\right\rceil 
\Biggr).
\]
In order to respect the target finality \(T_{\text{finality}}\), we then cap the bundle epoch so that super proving, batching, and bundling all fit within the finality limit:
\begin{multline}
\text{Epoch}_{\text{bundle}} = 
\min\!\Biggl(
\max\!\Bigl(
\text{Epoch}_{\text{batch}},\;\\
\left\lceil 
\dfrac{
\text{Max}_{\text{batch proofs per bundle}}
\,\cdot\,
\text{Epoch}_{\text{batch}}
}{
N_{\text{batch}}
} 
\right\rceil
\Bigr),
\\
T_{\text{finality}} 
-\,
\text{Epoch}_{\text{batch}} 
-\,
T_{\text{super}}
\Biggr).
\end{multline}

\subsubsection{Number of Batches Per Bundle Epoch}
The number of batch proofs that appear in each bundle epoch is determined by:
\begin{equation}
N_{\text{batches per bundle epoch}} 
= 
\dfrac{
\text{Epoch}_{\text{bundle}}
}{
\text{Epoch}_{\text{batch}}
}.
\end{equation}

\subsubsection{Number of Bundle Provers}
The number of bundle provers is determined by the number of batch proofs generated in the bundle epoch and the bundling capacity of a single prover. Specifically, each bundle prover can handle \(\left\lfloor \frac{\text{Bundle Epoch}}{T_{\text{bundle}}} \right\rfloor\) bundles within one epoch, with each bundle aggregating up to \(\text{Max}_{\text{batch proofs per bundle}}\) batch proofs:

\begin{equation}
N_{\text{bundle}} 
= 
\left\lceil 
\dfrac{
N_{\text{batch proofs per epoch}}
}{
\text{Max}_{\text{batch proofs per bundle}}
\,\cdot\,
\left\lfloor 
\dfrac{
\text{Bundle Epoch}
}{
T_{\text{bundle}}
}
\right\rfloor
}
\right\rceil.
\end{equation}

\paragraph*{Explanation}
\begin{itemize}
    \item \(N_{\text{batch proofs per epoch}}\): The total number of batch proofs generated during one bundle epoch.
    \item \(\text{Max}_{\text{batch proofs per bundle}}\): The maximum number of batch proofs that can be processed in a single bundle.
    \item \(\text{Bundle Epoch}\): The time duration of one bundle epoch.
    \item \(T_{\text{bundle}}\): The time it takes for a single bundle prover to process one bundle.
    \item \(\left\lfloor \frac{\text{Bundle Epoch}}{T_{\text{bundle}}} \right\rfloor\): The number of bundles a single prover can process in one epoch.
\end{itemize}

\subsubsection{Costs}
Data availability (DA) and verification costs can be modeled in two ways: \emph{blob-based} or \emph{calldata-based}.

\paragraph{Blob-based Data Availability Cost Per Batch}
\begin{multline}
\text{Cost}_{\text{DA per batch}} = 
\text{Gas}_{\text{DA per batch}} 
\,\times\, 
\text{Price}_{\text{gas (gwei)}} 
\,\times\, 
10^{-9} \\
\,\times\, 
\text{Price}_{\text{ETH (USD)}}
\end{multline}

\paragraph{Calldata-based Data Availability Cost Per Batch}
\begin{multline}
\text{Cost}_{\text{DA per batch}} = 
\text{Length}_{\text{batch (bytes)}} 
\,\times\, 
\text{Gas}_{\text{DA per byte}}
\,\times\, \\
\text{Price}_{\text{gas (gwei)}}
\,\times\, 
10^{-9}
\,\times\, 
\text{Price}_{\text{ETH (USD)}}
\end{multline}

\paragraph{Verification Cost Per Bundle}
\begin{multline}
\text{Cost}_{\text{verification per bundle}} 
= 
\text{Gas}_{\text{verification per bundle}}
\,\times\,
\text{Price}_{\text{gas (gwei)}}\\
\,\times\,
10^{-9}
\,\times\,
\text{Price}_{\text{ETH (USD)}}
\end{multline}

\paragraph{L1 Costs}
\begin{align}
\text{Cost}_{\text{DA per batch epoch}} 
&= 
N_{\text{batch}} 
\,\times\, 
\text{Cost}_{\text{DA per batch}}, 
\\
\text{Cost}_{\text{verification per bundle epoch}} 
&= 
N_{\text{bundle}}
\,\times\,
\text{Cost}_{\text{verification per bundle}}
\end{align}

\begin{multline}
\text{Cost}_{\text{L1 per bundle epoch}} 
= 
\sum_{i=1}^{N_{\text{batches per bundle epoch}}} 
\text{Cost}_{\text{DA per batch}}
\;+\;\\
\text{Cost}_{\text{verification per bundle}}.
\end{multline}

\paragraph{Machine Costs}
\begin{multline}
\text{Cost}_{\text{Total Machine}} 
= 
\bigl( 
N_{\text{super}} 
\,\times\, 
\text{Cost}_{\text{super machine}} 
\bigr)
\;+\;\\
\bigl( 
N_{\text{batch}} 
\,\times\,
\text{Cost}_{\text{batch machine}} 
\bigr)
\\
+\;
\bigl( 
N_{\text{bundle}} 
\,\times\,
\text{Cost}_{\text{bundle machine}} 
\bigr).
\end{multline}

\subsection{Average expected DA size of Chunk Proofs}
Because we publish data availability (DA) at the batch level, the DA byte‐budget caps how many super‐circuit proofs each batch can contain. In our benchmarks we set
\[
  \text{Max}_{\text{super proofs per batch}} = 45,
\]
which approximates a mix of simple transfers and moderate‐size deployments. In real‐world workloads, heavier transactions (e.g.\ large smart‐contract creations) consume more DA bytes per proof, reducing this limit to around 15–20 proofs for the same batch.  

For production use, model users should first measure their median DA usage per proof and then set \(\text{Max}_{\text{super proofs per batch}}\) accordingly. This ensures the cost and capacity estimates accurately reflect their actual transaction mix.

\subsection{Z3 SMT Solver}\label{sec:design-z3}

We encode our sizing model directly in Z3’s Python API to minimize total monthly cost subject to TPS and finality constraints. Key steps:

\begin{enumerate}
  \item \textbf{Optimize context}: use \verb|opt = Optimize()| with tuned parameters:
    \begin{verbatim}
set_param('auto_config', True)
set_param('parallel.enable', True)
z3.set_param('parallel.threads.max', 64)
    \end{verbatim}
  \item \textbf{Variables}  
    \begin{itemize}
      \item \emph{Inputs} (Real/Int): TPS, circuit times (\texttt{time\_super}, \texttt{time\_batch}, \texttt{time\_bundle}), gas rates, machine costs, \texttt{target\_finality}.
      \item \emph{Decisions} (Int): \texttt{num\_super\_provers}, \texttt{num\_batch\_provers}, \texttt{num\_bundle\_provers}.  
      \item \emph{Epochs} (Real): \texttt{batch\_epoch}, \texttt{bundle\_epoch}.
    \end{itemize}
  \item \textbf{Constraints}  
    \begin{itemize}
      \item \emph{Capacity}—ensure each layer’s prover count can handle its incoming proofs:
      \[
        N_\ell \times \bigl(\tfrac{\text{epoch}_\ell}{T_\ell}\times \text{max\_proofs}_\ell\bigr)
        \ge
        \text{demand}_\ell
      \]
      \item \emph{Finality}—enforce
      \[
        T_{\text{super}} + \mathit{batch\_epoch} + \mathit{bundle\_epoch}
        \;\le\;\mathit{target\_finality}.
      \]
      \item \emph{Ordering}—batch starts after super, bundle after batch.
    \end{itemize}
    \item \textbf{Cost objective}  
    \[
    \begin{split}
    \mathit{total\_cost} &= \sum_{\ell\in\{\mathrm{super,batch,bundle}\}} N_\ell\,C_\ell \\ 
    &\quad+ (\mathit{monthly\_batches}\times \mathrm{DA\_USD}) \\ 
    &\quad+ (\mathit{monthly\_bundles}\times \mathrm{Verif\_USD})
    \end{split}
    \]
    and call \verb|opt.minimize(total_cost)|.  
 \item \textbf{Solve and extract decisions}%

\end{enumerate}

This implementation automatically yields the integer prover counts and epoch lengths that satisfy both throughput and finality at minimum cost.

\subsection{Simulator}
\label{subsec:simulator-design}

We implement a discrete-event simulator to model the behavior of a proving pipeline consisting of super provers, batch provers, and bundle provers. The simulator is designed to detect lag conditions under resource constraints and estimate associated costs across epochal time steps. The design closely mirrors the logic in Algorithm~\ref{alg:sim-loop}.

\subsubsection{Architecture Overview}

The simulator operates over two interleaved epochs: \emph{batch epochs} and \emph{bundle epochs}. Super provers continuously generate proofs, which are then consumed by batch provers. The outputs of batch provers are in turn grouped into bundles. Each prover tier is constrained by its capacity and time bounds, and lag is detected when a pool exceeds downstream capacity.

\begin{figure}[H]
\centering
\begin{tikzpicture}[
  node distance=.5cm and 1.8cm,
  every node/.style={font=\scriptsize, align=center}
]

  % Pools
  \node (proofpool)  [draw, rectangle, minimum width=2cm, minimum height=0.6cm] {Super Proof Pool\\(\texttt{proof\_pool})};
  \node (batchpool)  [draw, rectangle, below=of proofpool] {Batch Proof Pool\\(\texttt{batch\_pool})};
  \node (finalbundles) [draw, rectangle, below=of batchpool] {Final Bundles};

  % Provers
  \node (batchprover)  [draw, rectangle, right=of proofpool, minimum width=1.8cm] {Batch\\Provers};
  \node (bundleprover) [draw, rectangle, right=of batchpool, minimum width=1.8cm] {Bundle\\Provers};

  % Super prover source
  \node (superprover) [draw, rectangle, left=of proofpool, minimum width=.8cm] {Super\\Provers};

  % Arrows
  \draw[->, thick] (superprover.east) -- (proofpool.west) node[midway, above] {\scriptsize generate};
  \draw[->, thick] (proofpool.east) -- (batchprover.west) node[midway, above] {\scriptsize consume};
  \draw[->, thick] (batchprover.south) -- (batchpool.east) node[midway, right] {\scriptsize produce};

  \draw[->, thick] (batchpool.east) -- (bundleprover.west) node[midway, above] {\scriptsize consume};
  \draw[->, thick] (bundleprover.south) -- (finalbundles.east) node[midway, right] {\scriptsize produce};

\end{tikzpicture}
\caption{Data and control flow in the simulation: provers consume from pools and populate downstream pools.}
\label{fig:sim-architecture}
\end{figure}
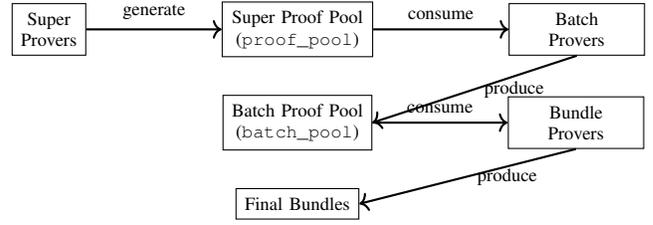

\subsubsection{Simulation Flow}

The simulation proceeds by selecting the next closest event—either a batch epoch or a bundle epoch—accumulating proofs up to that point, and executing the corresponding process. Accumulated proofs may be fractional if time between events does not align with the prover interval.

\begin{algorithm}[H]
\caption{Main Simulation Loop}
\label{alg:sim-loop}
\begin{algorithmic}[1]
\REQUIRE Parameters \texttt{params}, derived values \texttt{derived}, max bundle epochs $M$
\STATE Initialize: \texttt{simulation\_time} $\gets 0$, counters $\gets 0$, pools $\gets 0$
\WHILE{\texttt{bundle\_epoch\_passed} $< M$}
    \STATE $t_\text{batch} \gets (e_\text{batch} + 1) \cdot T_\text{batch}$
    \STATE $t_\text{bundle} \gets (e_\text{bundle} + 1) \cdot T_\text{bundle}$
    \STATE $t_\text{next} \gets \min(t_\text{batch}, t_\text{bundle})$
    \STATE \texttt{accumulate\_super\_proofs($t_\text{next}$)}
    \IF{$t_\text{next} == t_\text{batch}$}
        \STATE $e_\text{batch} \gets e_\text{batch} + 1$
        \IF{\texttt{proof\_pool} exceeds batch capacity}
            \STATE Record \texttt{batch prover lag}; \textbf{break}
        \ENDIF
        \STATE Consume \texttt{proof\_pool}, update \texttt{batch\_pool}
    \ENDIF
    \IF{$t_\text{next} == t_\text{bundle}$}
        \STATE $e_\text{bundle} \gets e_\text{bundle} + 1$
        \IF{\texttt{batch\_pool} exceeds bundle capacity}
            \STATE Record \texttt{bundle prover lag}; \textbf{break}
        \ENDIF
        \STATE Consume \texttt{batch\_pool}, finalize bundles
    \ENDIF
\ENDWHILE
\STATE Return final \texttt{SimulationResults}
\end{algorithmic}
\end{algorithm}

\subsubsection{Lag Detection}

The simulator reports a prover lag if resource pools exceed their respective consumption limits:

\paragraph{Batch Prover Lag:}
\begin{equation}
|\text{Super proof pool}| > N_{\text{batch}} \cdot \text{Max}_{\text{super/batch}}
\end{equation}

\paragraph{Bundle Prover Lag:}
\begin{multline}
|\text{Batch proof pool}| > \text{Max}_{\text{bundle proofs per epoch}} \cdot \\
\text{Max}_{\text{batch proofs per bundle}}
\end{multline}

\subsubsection{Transaction Throughput Guard}

To ensure that super prover capacity can handle the transaction load, we enforce the inequality:
\begin{equation}
\text{TPS} \cdot T_{\text{super}} \leq N_{\text{super}} \cdot \text{Tx}_{\max}
\end{equation}

Violations of this condition indicate underprovisioning and require scaling the proving layer.

\section{Evaluation}
\textbf{Objectives.}
We assess the cost model on three axes:  
(i)~\emph{Correctness} – the roll-up always meets the target finality;  
(ii) \emph{Liveness} – no backlog accumulates under observed peak loads;  
(iii) \emph{Cost} – monthly \$-expenditure is minimised.
The section proceeds as follows: experimental set-up (§\ref{sec:eval-setup}); a naïve baseline (§\ref{sec:eval-baseline-cost}); optimised sizing via \textsc{Z3} (§\ref{sec:eval-z3}); and validation with a discrete-event simulator (§\ref{sec:eval-sim}). %and external validity on public L2s (§\ref{sec:eval-external}). %; and threats to validity (§\ref{sec:eval-threats}).

% ---------------------------------------------------------
\subsection{Experimental Set-up}
\label{sec:eval-setup}

\subsubsection{Hardware candidates}
Table~\ref{tab:memory_usage_gb} reports the memory upper bounds of each prover step.
Only machines able to satisfy the most demanding step (\emph{Super Prover}, 140.6\,GiB)
were retained for benchmarking.

\begin{table}[htbp]
  \centering
  \caption{Peak resident memory per proving step.}
\label{tab:memory_usage_gb}
\begin{center}
\begin{tabular}{|c|c|c|}
\hline
\textbf{Proving Step} & \textbf{Circuit} & \makecell{\textbf{Memory} \\ \textbf{Usage (GB)}} \\ \hline
Super & super-circuit & 140.65 \\ \hline
% Super Compress Wide & compress-poseidon-wide & 63.07 \\ \hline
% Super Compress Thin & compress-poseidon-thin & 48.35 \\ \hline
Batch & batch-circuit & 138.37 \\ \hline
% Batch Compress Thin & compress-poseidon-thin & 27.85 \\ \hline
Bundle & bundle-circuit & 48.30 \\ \hline
% Bundle Prover & bundle-circuit & 21.68 \\ \hline
% Bundle Compress Thin & compress-keccak-thin & 48.30 \\ \hline
\end{tabular}
\end{center}
\end{table}

\subsubsection{Chosen configurations}
Table~\ref{tab:pricing_performance} lists the two machines used throughout the study,
together with their monthly cost and measured runtimes.\footnote{
  Prices: 2025-04 USD list for AWS m7i.24xlarge and an in-house Metal RockawayX node.
}
% \subsubsection{Runtime on Various Machines}
% \autoref{tab:pricing_performance} provides pricing and proof computation benchmarks for the selected machines. This table includes details such as monthly costs, backend configurations, and execution times for each proving step.
\begin{table}[t!]
\caption{Pricing and Proof Computation Benchmark Per Machine. The prices are the monthly fee for retaining the machines at the time of writing. }
\label{tab:pricing_performance}
\begin{center}
 \resizebox{\columnwidth}{!}{
\begin{tabular}{|c|c|c|c|c|c|}
\hline
\textbf{Config.} & \textbf{Price}  & \textbf{Chunk} & \textbf{Batch} & \textbf{Bundle} \\ \hline
 GPU: 8x NVIDIA L4 24GB & \$1,000  & 1592.81s & 1718.2s & 518s
\\ \hline
 CPU: m7i.24xlarge (96C) & \$2,839 & 2352s & 1952s & 804s \\ \hline
\end{tabular}
}
\end{center}
\end{table}

% c7i.48xlarge & 5,668 & Halo2 CPU & 2108 & 1976 & 0 \\ \hline
% z1d.12xlarge & 2,812 & Halo2 CPU & 4436 & 3756 & 0 \\ \hline
% c7a.24xlarge & 3,259 & Halo2 CPU & 2250 & 2084 & 0 \\ \hline
% g2-standard-96 (8x L4 GPU) & 5,889 & Icicle V2 GPU & 3120 & 2604 & 0 \\ \hline
% g2-standard-96 (8x L4 GPU) & 5,889 & Icicle V3 GPU & 2234 & 0 & 0 \\ \hline
% Metal (Dell) & 17300 & Halo2 CPU & 4020 & 4103 & 59 \\ \hline
% Metal Rockawayx Halo2 CPU & 1,000 & Halo2 CPU & 2654.682 & 2863.652 & 0 \\ \hline

% Next, we need to choose a realistic finality target that aligns with industry practices.

% ---------------------------------------------------------
\subsection{Baseline Cost Analysis}
\label{sec:eval-baseline-cost}

\paragraph{Sizing policy.}
The baseline dynamically allocates provers as follows:

\begin{itemize}
  \item \textbf{Chunk provers} scale linearly with throughput:
        \(N_{\text{chunk}} \propto \text{TPS}\).
  \item \textbf{Batch provers} are provisioned at one per
        \(45\) chunk provers.
  \item \textbf{Bundle provers} — exactly one, as bundling is much faster.
  \item \textbf{Finality guard} — fleets are scaled until
        \(T_{\text{finality}} \le 14\,400\;\text{s}\;(4\text{ h})\).
\end{itemize}

\paragraph{Stress events.}
We captured the TPS and blob-fee for the four events that listed in Table~\ref{tab:baseline-events}). and fed into the sizing rules above.

\begin{table}[htbp]
  \centering
  \caption{Real-world events used as baseline workloads.}
  \label{tab:baseline-events}
  \begin{tabularx}{\linewidth}{@{}lX@{}}
    \toprule
    \textbf{Event} & \textbf{Observed impact} \\ \midrule
    \textbf{Steady} - steady and consistent traffic & TPS is consistently around 0.10. \\
    \textbf{TGE} - Zircuit token launch (2024-11-18) & TPS surged from 0.04 to 2.12. \\
    % Ethereum congestion (2025-03-11)  & TPS peak 0.54 (×5 over median). \\
    \textbf{Blob fee spike}   & Blob base fee rose from
      \$4.7×10$^{-9}$ to \$3450. \cite{blobCostDune} \\ 
    \textbf{Surge} - high traffic on Zircuit & TPS can surge to 100. \\

      \bottomrule
  \end{tabularx}
\end{table}

\paragraph{Cost results.}
We use the hardware costs and proving times from Table~\ref{tab:pricing_performance}, together with the scenario TPS, fees, and the other model parameters listed in Section~\ref{sec:fixed-params}. The cost model indicates that GPU hardware is the least expensive, so we select it. Table~\ref{tab:baseline-cost} shows the breakdown of monthly operational expenses (DA cost, verification cost, machine cost, and total monthly cost in USD) for each of the four scenarios.
\begin{table}[t!]
  \centering
  \caption{Naïve dynamic sizing metrics by scenario.}
  \label{tab:baseline-cost}
  \resizebox{\columnwidth}{!}{%
    \begin{tabular}{@{}l c c c c c c@{}}
      \toprule
      Scenario & TPS & \shortstack{Chunk/\\Batch/\\Bundle} & \shortstack{DA\\Cost} 
               & \shortstack{Verification\\Cost} & \shortstack{Machine\\fee} 
               & Total \\
      \midrule
      Steady      & 0.10  & 2/1/1     & \$27\,350.2    & \$42\,638.9    & \$4\,000      & \$73\,989.1    \\
      TGE         & 2.12  & 27/1/1    & \$27\,350.2    & \$42\,638.9    & \$29\,000     & \$98\,989.1    \\
      Blob Spike  & 0.10  & 2/1/1     & \$7\,548\,660  & \$42\,638.9    & \$4\,000      & \$7\,591\,298.9  \\
      Surge       & 100   & 1\,245/30/2 & \$820\,507   & \$85\,277.9    & \$1\,277\,000  & \$2\,182\,780    \\
      \bottomrule
    \end{tabular}%
  }
\end{table}
% ---------------------------------------------------------
\subsection{Optimised Sizing via \textsc{Z3}}
\label{sec:eval-z3}

Using the same workloads, we encode the non-linear constraints from § \ref{sec:design-z3} in \textsc{Z3}, minimizing total monthly cost. We set the solver’s timeout to 6 000 seconds. Table \ref{tab:z3-optimized-benchmark} presents the results of the \textsc{Z3}-based optimization. The model leverages two delay parameters—batch and bundle epochs—to fully utilize circuit capacity, avoiding the suboptimal proofs observed in the four experimental scenarios.
% \FloatBarrier
\begin{table*}[t!]
  \centering
  \caption{Optimised fleet per scenario (machine chosen by lowest total \$).}
  \label{tab:z3-optimized-benchmark}
  \begin{tabular}{@{}lcccccccccc@{}}
    \toprule
    \textbf{Scenario} 
      & \textbf{TPS} 
      & \textbf{\shortstack{Chunk/\\Batch/\\Bundle}} 
      & \textbf{\shortstack{Batch\\epoch}} 
      & \textbf{\shortstack{DA\\cost}} 
      & \textbf{\shortstack{Bundle\\epoch}} 
      & \textbf{\shortstack{Verification\\cost}} 
      & \textbf{\shortstack{Machines\\fee}} 
      & \textbf{\shortstack{Total\\before}} 
      & \textbf{\shortstack{Total\\after}} 
      & \textbf{Reduction} \\ 
    \midrule
    Steady           & 0.10 & 2/1/1     & 6337s & \$7,400       & 6337s & \$11,000    & \$4,000     & \$73,989.1      & \$22,000      & 70\% \\
    TGE              & 2.12 & 27/1/1    & 6337s & \$7,400       & 6337s & \$11,000    & \$29,000    & \$98,989.1      & \$47,000      & 53\% \\
    Blob fee Spike   & 0.10 & 2/1/1     & 5799s & \$2,200,000   & 7005s & \$10,000    & \$4,000     & \$7,591,298.9   & \$2,200,000   & 71\% \\
    Surge            & 100  & 1245/30/2 & 5297s & \$267,000     & 7491s & \$9,000     & \$1,280,000 & \$2,182,780     & \$1,500,000   & 31\% \\
    \bottomrule
  \end{tabular}
\end{table*}
% \FloatBarrier

Relative to the naïve fleet (Table~\ref{tab:baseline-cost}),
\textsc{Z3} cuts machine cost by up to 71\% while preserving the expected target finality of 4 hours.

% ---------------------------------------------------------
\subsection{Validation by Simulation}
\label{sec:eval-sim}
To validate that the Z3 solution incurs no lags and meets the expected finality (as defined in our simulator, § \ref{subsec:simulator-design}), we replayed 30 days of each scenario using the optimized fleet configurations. No batch- or bundle-pool backlog was observed, and end-to-end finality never exceeded 14 400 s.
\section{Limitations}

\subsection{Data Availability Size and Batcher Limit}
We assumed that the data availability (DA) passed to the batches remains fixed and can always fit within a batch, even when transactions have excessive DA usage. In practice, this assumption may not hold. The batcher is further constrained by the total size of the DA associated with the super proofs, limiting the number of super proofs it can include. Consequently, our cost estimation serves as a middle-ground, first-order approximation rather than a precise calculation.

\section{Conclusions}
In this paper, we presented a comprehensive model to capture the costs associated with ZK-rollups and outlined all relevant parameters from a widely used zkEVM Halo2 ZK-rollup configuration. We developed a simulator to evaluate our model and demonstrated how it effectively prevents system lags while building a cost-efficient configuration. Additionally, we implemented our system in the Z3 solver to search for cost-optimal solutions under various constraints, highlighting the promise of automated reasoning for handling non-linear aspects of our model. 

\textbf{Future work} could explore an architecture to dynamically based on the TPS and the Z3 model optimizes costs on the go.  Another direction is to factor in the DA size as a parameter in the model. Both avenues hold promise for pushing the scalability and efficiency of ZK-rollup designs even further.    

    %% Bibliography
	\if\Class0
	    \bibliographystyle{splncs04}
	\else
	    \bibliographystyle{alpha}
	\fi
    \bibliography{bib/custom}

    %% Appendixes    
    %\appendix
    %\input{tex/appendix.tex}
    
\end{document}